\documentclass{elsart} 
\usepackage{epsfig}  
\usepackage{amssymb}  
\usepackage[longnamesfirst]{natbib}  

\def \bu{ {{\mbox {\boldmath $1$}}}   }
\def \bs{ {{\mbox {\boldmath $\sigma \! $}}}}
\def \btau{ {{\mbox {\boldmath $\tau$}}}}
\def \bgam{ {{\mbox {\boldmath $\gamma$}}}}

\def \br{ {{\mbox {\boldmath $r$}}}}

\def \bF{ {{\mbox {\boldmath $F$}}}}

\def \bQ{ {{\mbox {\boldmath $Q$}}}}

\def \bY{ {{\mbox {\boldmath $Y$}}}}
\def \bZ{ {{\mbox {\boldmath $Z$}}}}

\def \bsh{ {\hat \bs}}
\def \bk{ {\mbox {\boldmath $\kappa$}}}
\def \avel { {\bigl\langle}}
\def \aver { {\bigr\rangle}}

\def \bDel{ {\bf \Delta}}
\def \bPi{ {\bf \Pi}}

\def \one{ { { 1 \kern-0.28em {\rm I}}  }}
\def \bsjk{ {\bs_{jk}}}

\begin{document} 

\begin{frontmatter}
 
\title{ {The influence of the extent of excluded volume interactions  
on the linear viscoelastic properties of dilute polymer solutions }} 
\author {{J. Ravi Prakash}\thanksref{Presentaddress}}
\address{ Department~of~Chemical~Engineering,
Indian~Institute~of~Technology, Madras, India 600 036}  
\thanks[Presentaddress]{Present address:  
Department~of~Chemical~Engineering, Monash University,
Victoria 3800, Australia }
 
\begin{abstract} 
		The Rouse model has recently been modified to take
		into account the excluded volume interactions that
		exist between various parts of a polymer chain by
		incorporating a narrow Gaussian repulsive potential
		between pairs of beads on the Rouse
		chain~\citep*{prakev2000}. The narrow Gaussian
		potential is characterized by two parameters: $z^*$
		--- which accounts for the {\em strength\/} of the
		interaction, and $d^*$ --- which accounts for the {\em
		extent\/} of the interaction. In the limit of $d^*$
		going to zero, the narrow Gaussian potential tends to
		the more commonly used $\delta$-function repulsive
		potential. The influence of the parameter $d^*$, in
		the limit of infinite chain length, on equilibrium and
		linear viscoelastic properties, and on universal
		ratios involving these properties, is examined here. A
		renormalization group calculation of the end-to-end
		vector suggests that the value chosen for the variable
		$d^*$ will not affect critical exponents, or universal
		ratios. A similar trend is also observed for results
		obtained with an approximate solution, which is based
		on the assumption that the non-equilibrium
		configurational distribution function is Gaussian.
\end{abstract} 


\end{frontmatter} 

\section{Introduction}

The macroscopic consequences, on both equilibrium and rheological
properties of dilute polymer solutions, of the microscopic fact that
two parts of a polymer chain cannot occupy the same place at the same
time, have been the subject of innumerable investigations. Analytical
treatments of this {\em excluded volume\/} effect have, by and large,
modeled the repulsive interaction between parts of the polymer chain
with a Dirac delta potential. As a result, different parts of the
chain interact with each other only when they are directly in contact
\citep{yama71,doi86,ottrg89,declos90,zylkarg91,schaf99}. The use of a
$\delta$-function potential makes it difficult to obtain {\em exact\/}
analytical results.  Consequently, progress has usually been made by
using renormalization group methods to refine the results of
perturbation calculations. On the other hand, numerical investigations
with Monte Carlo and Brownian dynamics simulations---which lead to
exact results---have been based on potentials that have a finite range
of excluded volume interaction, such as the Lennard-Jones potential,
because of the unsuitability of the $\delta$-function potential for
numerical simulations~\citep{gras99,cifre99}.  The quality of the
approximate analytical results are usually assessed---at least in the
case of static properties, which have been much more extensively
studied than properties away from equilibrium---by comparison with the
exact Monte Carlo simulations. The comparison is made, in spite of the
fact that the analytical and numerical approaches are based on
different potentials, because it is generally accepted that the choice
of the excluded volume potential does not influence the scaling of
observables in the limit of long chains. Such comparisons have
established that the results of static theories agree well with both
Monte Carlo simulations and with experimental
observations~\citep{schaf99}.

Prakash and \"Ottinger have recently examined the influence of
excluded volume effects on the rheological properties of dilute
polymer solutions by representing the polymer molecule with a Hookean
dumbbell model, and using a narrow Gaussian repulsive potential to
describe the excluded volume interactions between the beads of the
dumbbell~\citep{prakottev99}. The same potential has subsequently been
used by Prakash~\citep{prakev2000} to represent the excluded volume
interactions between pairs of beads in a bead-spring chain model for
the polymer molecule. If the bead position vectors ${\br}_{\nu}, \,
\nu = 1,2, \ldots, N,$ are used to denote the instantaneous
configuration of the bead-spring chain (which consists of $N$ beads
connected together by $(N-1)$ Hookean springs), the narrow Gaussian
potential $ E \left( {\br}_{\nu} - {\br}_{\mu} \right)$ between any
pair of beads $\nu$ and $\mu$ can be written as,
\begin{equation}
E \left( {\br}_{\nu} - {\br}_{\mu} \right) = 
\left( {z^* \over {d^*}^3} \right) k_{\rm B} T 
\, \exp \left\lbrace - {H \over 2 k_{\rm B} T }\, 
{ \br_{\nu \mu}^2 \over {d^*}^2} \right\rbrace
\label{evpot}
\end{equation}
where, $ k_{\rm B}$ is Boltzmann's constant, $T$ is the absolute
temperature, $H$ is the spring constant, $\br_{\nu \mu} =
\br_{\nu}-\br_{\mu}$, is the vector between beads $\mu$ and $\nu$, and
the parameters $z^*$ and $d^*$ are non-dimensional quantities which
characterize the narrow Gaussian potential.  While $z^*$ measures the
strength of the excluded volume interaction, $d^*$ is a measure of the
extent of excluded volume interaction.  Note that in the limit $d^*$
tending to zero, the narrow Gaussian potential becomes a
$\delta$-function potential. Compared to the $\delta$-function 
potential, analytical calculations are not significantly harder with 
the narrow Gaussian potential; often, upon setting $d^*=0$, the
predictions of a $\delta$-function potential  
can be obtained. Furthermore, Brownian dynamics simulations 
can be performed with a narrow Gaussian potential. As a result, the 
use of a narrow Gaussian potential makes it possible to compare
approximate analytical results and numerical simulations consistently. 

Two approximate solutions have been developed in order to
examine the consequences of using a narrow Gaussian
potential~\citep{prakev2000}. The first approximation is a first order
perturbation expansion in the strength of excluded volume interaction, 
while the second approximation, which is non-perturbative
in nature, is based on assuming that the configurational
distribution function is a Gaussian distribution. The
equilibrium swelling of the end-to-end vector and the radius of
gyration, and zero shear rate properties, predicted by both
the approximate solutions, were compared with the predictions of exact 
Brownian dynamics simulations.

The Brownian dynamics simulations revealed, rather unexpectedly, that
for chains with an {\em arbitrary} but {\em finite} number of beads,
the use of a $\delta$-function excluded volume potential leads to
predictions that are identical to those of the Rouse model. In other
words {\em theta} and {\em good} solvents were predicted to have the
same equilibrium and rheological behavior. Departure from Rouse model
predictions could only be obtained for non-zero values of $d^*$. As
the number of beads $N$ became large, however, the simulation results
seemed to suggest that the use of a $\delta$-function potential may be
justified. It must be noted though that this conclusion is
speculative, and is based on rather limited data. It is clearly
worthwhile therefore to examine the equilibrium and rheological
predictions obtained by carrying out the two limits $d^* \to 0$ and $N
\to \infty$ simultaneously. However, it is difficult to obtain data
for very long chains because of the computational intensity of
Brownian dynamics simulations.

In contrast to Brownian dynamics simulations, both the first order
perturbation expansion and the Gaussian approximation predict a
spurious non-trivial effect due to the presence of $\delta$-function
excluded volume interactions, i.e. when $d^* = 0$.  In the case of a
dumbbell model it was shown rigorously that the source of the problem
was the incorrect term-by-term integration of a series that was not
uniformly convergent~\citep{prakottev99}. On the other hand, for fixed
values of $N$ and $z^*$, the Gaussian approximation becomes
increasingly accurate as $d^*$ increases. Furthermore, for fixed
values of the re-scaled variables $d = d^* / \sqrt N$, and $z= {z^*}
\sqrt N$, the accuracy of the Gaussian approximation increases as $N$
increases~\citep{prakev2000}.  Since the Gaussian approximation is not
as computationally intensive as Brownian dynamics simulations, it is
possible to obtain predictions in the long chain limit by accumulating
data for chains of finite length and extrapolating to $N \to
\infty$. The Gaussian approximation provides a means therefore of
exploring the role of the extent of excluded volume interaction in the
limit of long chains. The purpose of this paper is to use the
equilibrium and zero shear rate predictions of the Gaussian
approximation to gain insight into the influence of the parameter $d^*$
as $N \to \infty$.

As a preliminary motivation to examining the Gaussian approximation in
the limit of long chains, an elementary renormalization group
calculation of the equilibrium end-to-end vector is also
carried out in this paper. The renormalization group method is a
systematic way of examining the macroscopic {\it relevance} of a
microscopic parameter, i.e., if it has macroscopic consequences. It is
therefore an ideal, albeit technically involved, tool to examine the
role of the parameter $d^*$. 

The plan of the paper is as follows. The theoretical background for
the Gaussian approximation, and the principal results obtained on
making the Gaussian approximation are summarized in 
the next section. The renormalization of the equilibrium end-to-end
vector is taken up in section~3. The results of extrapolating 
finite chain Gaussian approximation results to infinite chain length
are discussed in section~4, and the main conclusions of the present
work are summarized in section~5. 

\section{The Gaussian approximation}

Two fundamental equations, (i) the {\em diffusion} equation for
the configurational distribution function of the polymer chain, and
(ii) the {\em Kramers} expression for the polymer contribution to the
stress tensor, are the basic ingredients of a kinetic theory of
dilute polymer solutions. In the presence of excluded volume
interactions, one can show that the diffusion equation for a
bead-spring chain with $N$ beads, suspended in a Newtonian solvent,
has the form~\citep{prakev2000},  
\begin{eqnarray}
{\partial \, \psi \over \partial t} = &-& \sum_{j=1}^{N-1} \,
{\partial   \over \partial {\bQ}_j} \cdot \biggl( 
\bk \cdot {\bQ}_{j} 
- {H \over \zeta} \, \sum_{k=1}^{N-1} \, A_{jk} \, {\bQ}_k
+ {1 \over \zeta} \, \sum_{\nu=1}^{N}\, {\overline B}_{j \nu} \bF_{\nu}^{(E)}
\, \biggr) \, \psi \nonumber \\
\nonumber \\
&+& {k_{\rm B} T \over \zeta} \, \sum_{j,\, k=1}^{N-1} \, A_{jk} \; 
{\partial \over \partial {\bQ}_j} \cdot
{\partial \psi \over \partial {\bQ}_k} 
\label{diff}
\end{eqnarray}
where, ${\bQ}_i= {\br}_{i+1} - {\br}_{i}$, is the {\it bead connector}
vector between the beads $i$ and $i+1$, $ \psi \, ({\bQ}_1, \ldots,
{\bQ}_{N-1}, t)$ is the configurational distribution function, $\bk
(t)$ is the traceless transpose of the velocity-gradient tensor,
$\zeta$ is the bead friction coefficient, ${\overline B}_{k \nu}$ is
an $(N-1) \times N$ matrix defined by, ${\overline B_{k \nu}} =
\delta_{k+1, \; \nu} - \delta_{k \nu}$, with $\delta_{k \nu}$ denoting
the Kronecker delta, and $A_{jk}$ is the Rouse matrix,
\begin{equation}
A_{jk}=\sum_{\nu=1}^{N}\, {\overline B}_{j \nu} {\overline B}_{k \nu} 
=\cases{ 2&for $\vert {j-k} \vert = 0 $,\cr
\noalign{\vskip3pt}
-1& for $\vert {j-k} \vert =1 $,\cr
\noalign{\vskip3pt}
0 & otherwise \cr} 
\label{rmatrix}
\end{equation}
The vector ${\bF}_{\nu}^{(E)}$ is the total {\em excluded volume} force on
bead $\nu$. It is given, in terms of the excluded volume potential
between the beads of the chain, by the expression,
\begin{equation}
{\bF}_{\nu}^{(E)} = - \sum_{\mu = 1 \atop \mu \ne \nu}^N \, {\partial 
\over \partial \br_\nu} \, E \left( {\br}_{\nu} - {\br}_{\mu} \right) 
\label{evforce}
\end{equation}

The Kramers expression for the polymer contribution to the stress
tensor, for a bead-spring chain model with Hookean springs and an
arbitrary excluded volume potential force, is given by~\citep{bird87b}, 
\begin{equation}
\btau^p = - n_{\rm p} H \,  \sum_{k=1}^{N-1} \, 
\avel {\bQ}_k \bQ_k \aver 
+ \bZ + (N-1) \,  n_{\rm p}  k_{\rm B} T \, \bu 
\label{kram}
\end{equation}
where, angular brackets denote an average performed with the
configurational distribution function $\psi$ (obtained by solving the
diffusion equation), $n_{\rm p}$ is the number density of polymers,
$\bu$ is the unit tensor, and the tensor $\bZ$, which represents the
{\em direct} contribution due to excluded volume
effects~\citep{prakev2000}, is given by, 
\begin{equation}
\bZ =  n_{\rm p} \, \sum_{\nu=1}^N 
\sum_{k=1}^{N-1} \, B_{\nu k} \, 
\avel \bQ_k {\bF}_{\nu}^{(E)} \aver 
\label{iso}
\end{equation}
The quantity $B_{\nu k}$ is a $N \times (N-1)$ matrix defined by, 
$B_{\nu k} = k/N - \Theta \, (k-\nu)$, with $\Theta \, (k-\nu)$
denoting a Heaviside step function.

The diffusion equation, Eq.~(\ref{diff}), becomes analytically
intractable when excluded volume interactions are described in terms
of the narrow Gaussian potential, Eq.~(\ref{evpot}). As mentioned in
the introduction, there are several
different solution schemes, involving either (i) an exact numerical
approach, or (ii) approximate analytical approaches, that can be
developed to overcome this problem. These have been discussed in
detail in~\citet{prakev2000}. Here, we summarize one approximate
non-perturbative solution procedure, called the Gaussian
approximation.

The Gaussian approximation is based on reducing the complex higher
order moments in Kramers expression, Eq.~(\ref{kram}), to functions of
only second order moments, by assuming that the non-equilibrium
configurational distribution function $\psi$, is a Gaussian
distribution. The Kramers expression can then be
shown to assume the form,
\begin{equation}
\btau^p = - n_{\rm p} H \,  \sum_{k=1}^{N-1} \, \bs_{kk} 
+ \bZ + (N-1) \,  n_{\rm p}  k_{\rm B} T \, \bu 
\label{gausskram}
\end{equation}
where, the $(N-1) \times (N-1)$ matrix of tensor components, 
$\bsjk = \avel {\bQ}_j{\bQ}_k \aver$, is the covariance matrix which
uniquely characterizes the Gaussian distribution, and 
the tensor $\bZ$ is now given by the expression,
\begin{equation}
\bZ =  {1 \over 2} \, z^* \, n_{\rm p}  k_{\rm B} T  
\sum_{\nu,  \, \mu=1 \atop \nu \ne \mu }^N  
\bsh_{\nu \mu} \cdot \bPi(\bsh_{\nu \mu})
\label{gaussiso}
\end{equation}
Here, the function $\bPi(\bsh_{\nu \mu})$ is given by,  
\begin{equation}
\bPi(\bsh_{\nu \mu}) = { \left[ {d^*}^2 \, \bu + \bsh_{\nu \mu} \right]^{-1}  
\over 
\sqrt{ \det \left( [ {d^*}^2 \, \bu + \bsh_{\nu \mu} ] \right) }}
\label{Pifun}
\end{equation}
with the tensors $\bsh_{\nu \mu}$ defined by,
\begin{equation}
\bsh_{\nu \mu} = \bsh_{\nu \mu}^T = \bsh_{\mu \nu} = {H \over k_B T }
\,\, \sum_{j,k \,= \,\min(\mu,\nu)}^{ \max(\mu,\nu)-1} \, \bsjk
\end{equation}

The Gaussian approximation is complete when a scheme for calculating
the covariance matrix $\bsjk$ is specified. A time evolution equation
for $\bsjk$ can be derived by multiplying the diffusion equation,
Eq.~(\ref{diff}), by ${\bQ}_j {\bQ}_k $ and integrating over all
configurations. The higher order moments that appear on the right hand
side of the evolution equation are reduced to second order moments by
consistently using the ansatz of the Gaussian approximation. The
following evolution equation for $\bsjk$ is then obtained,
\begin{eqnarray}
{d \over dt} \, \bsjk =
\bk \cdot \bsjk &+& \bsjk  \cdot \bk^T 
- { H \over \zeta} \, \sum_{m=1}^{N-1}  \left[ 
\bs_{jm} \, A_{mk} + A_{jm} \, \bs_{m k} \right] \nonumber \\ 
\nonumber \\ 
&+& { 2 k_{\rm B} T \over \zeta} \, A_{jk} \, \bu + \bY_{jk} 
\label{gaussecmom}
\end{eqnarray}
where, 
\begin{equation}
\bY_{jk} = {z^*} \, \left( { H \over \zeta} \right) \, \sum_{m=1}^{N-1}  
\left[ \bs_{jm} \cdot \bDel_{km} + \bDel_{jm} \cdot \bs_{m k} \right]
\label{gaussY}
\end{equation}
In Eq.~(\ref{gaussY}), the $(N-1) \times (N-1)$ matrix of tensor
components $\bDel_{jm}$ is defined by, 
\begin{equation}
\bDel_{jm} = \sum_{\mu=1}^{N}  \biggl\lbrace 
( B_{j+1,\, m} -  B_{\mu m}) \, \bPi (\bsh_{j+1,\, \mu})
- ( B_{j m} -  B_{\mu m}) \, \bPi (\bsh_{j \mu}) \biggr\rbrace 
\label{Del}
\end{equation}

The system of $9 \times (N-1)^2$ coupled ordinary differential
equations for $\bsjk$, Eq.~(\ref{gaussecmom}), can be solved
analytically (correct to first order in velocity gradient), by
expanding $\bsjk$ up to first order in velocity gradient about its
isotropic equilibrium value. On subsequently making use of Kramers
expression, Eq.~(\ref{gausskram}), the following first order
codeformational memory-integral expansion for the polymer contribution
to the stress tensor can be derived,
\begin{equation}
\btau^p= -   \int_{- \infty}^t d\!s \, G(t-s)\, 
{\bgam}_{[1]}(t,s) 
\label{memexp}
\end{equation}
where, ${\bgam}_{[1]}$ is the codeformational rate-of-strain
tensor~\citep{bird87a}, and $G(t)$ is the memory function whose
explicit form is given in appendix~B of~\citet{prakev2000}.

Exact expressions for the zero shear rate viscosity, $\eta_{p,0}$, and
the zero shear rate first normal stress difference, $\Psi_{1,0}$,
predicted by the Gaussian approximation, can be obtained from
Eq.~(\ref{memexp}). These expressions, which are lengthy, are given in
\citet{prakev2000}. It suffices here to note that 
their evaluation requires the inversion of an $(N-1)^2 \times (N-1)^2$
matrix. As a result, the CPU time required for their evaluation
scales as $N^6$.  Generating data for large values of $N$ becomes
extremely computationally intensive. Predictions of zero shear rate
properties have been obtained for chains up to a maximum of $N=40$
beads, since, for this value of $N$, a single run on an SGI Origin2000
computer with a 195 MHz processor required approximately 54 hours of
CPU time. 

It is appropriate now to summarize the most significant results
obtained on making the Gaussian approximation. When the equilibrium
swelling and the zero shear rate properties, at a constant value of
$z^*$, are plotted versus $d^*$ for various values of $N$, and
compared with Brownian dynamics simulations, it is found that, for
each value of $N$, the Gaussian approximation becomes accurate beyond
a threshold value of $d^*$. However, this threshold value increases as
$N$ increases, implying that the accuracy of the Gaussian
approximation decreases with increasing $N$. The picture changes
considerably, however, when the same data is viewed in terms of the
re-scaled extent of excluded volume interaction $d = d^* / \sqrt N$,
and re-scaled strength of the interaction $z= {z^*} \sqrt N$. The
Gaussian approximation appears in much better light in the context of
the re-scaled variables since, at a fixed value of $z$, it becomes
accurate over an increasingly larger range of values of $d$ as $N$
increases. Furthermore, as clarified below, asymptotic behavior in the
limit of large $N$ is observed.

In the limit of large $N$, curves for various values of $N$ begin to
collapse onto a single curve, indicating that all the equilibrium and
linear viscoelastic properties are independent of the value of $N$,
and are in fact, only functions of $z$ and $d$. This implies that, if
some knowledge about the leading order corrections to the infinite
chain length limit can be obtained, data accumulated for finite chains
for large enough values of $N$ can be efficiently extrapolated to the
limit $N \to \infty$. As will be seen in the section below, in
addition to elucidating the role of $d^*$ in the theory, a
renormalization group calculation of the equilibrium end-to-end vector
also provides insight into the leading order correction. Discussion of
the extrapolation procedure, and the results obtained on extrapolating
the Gaussian approximation data to infinite chains, will be taken up in
section~4.  

\section{Renormalization of the equilibrium end-to-end vector}

The second moment of the end-to-end vector 
$\br$ at equilibrium is given by the expression,
\begin{equation} 
\avel \br \br \aver_{\rm eq} =  \sum_{j, k=1}^{N-1}  
\avel \bQ_j \bQ_k \aver_{\rm eq}
\label{endtoend1}
\end{equation}
where, the suffix ``eq'' indicates that the average is carried out
with the equilibrium distribution function, $\psi_{\rm eq}$, given by, 
\begin{equation} 
\psi_{\rm eq} ({\bQ}_1, \ldots, {\bQ}_{N-1}) = {\mathcal N_{\rm eq}} 
\, e^{- {\phi / k_{\rm B} T}} 
\label{eqdist} 
\end{equation} 
The quantity $\phi$ denotes the potential energy of the bead-spring
chain, and ${\mathcal N_{\rm eq}}$ denotes the normalization constant.   
In the present model, the potential energy $\phi$ is the sum of the
potential energies of all the springs in the chain, and 
the excluded volume interaction energies between all pairs of beads
$\mu$ and~$\nu$, 
\begin{equation}
\phi = {1 \over 2} \, H \, \sum_{i=1}^{N-1} \, {\bQ}_i \cdot {\bQ}_i
+ {1 \over 2} \sum_{\mu,\nu = 1 \atop \mu \ne \nu}^N \, E \left
( {\br}_{\nu} - {\br}_{\mu} \right)
\label{poteng}
\end{equation}

In order to perform a renormalization group calculation of the
end-to-end vector, it is first necessary to expand the equilibrium
configurational distribution function in a perturbation expansion up
to first order in the strength of the excluded volume interaction. As
will be clear shortly, the expansion must be made in a space of
arbitrary dimensions $D$.

Upon expanding both the potential energy $\phi$ (Eq.~(\ref{poteng}),
with $E \left ( {\br}_{\nu} - {\br}_{\mu} \right)$ given by the narrow
Gaussian potential, Eq.~(\ref{evpot}), extended to $D$ dimensions),
and the normalization constant ${\mathcal N_{\rm eq}}$, to first order
in the strength of the excluded volume interaction, one can show that,
\begin{eqnarray}
\avel {\bQ}_j {\bQ}_k  \aver_{\rm eq} &=&
\left[ 1+ {1 \over 2 } \, z^*_D \sum_{\mu,\nu = 1 \atop \mu \ne \nu}^N
\, {1 \over \sqrt{ \det \left( {d^*}^2 \, \bu 
+ {(H / k_{\rm B} T )} \, \avel{\br}_{\nu \mu} 
{\br}_{\nu \mu} \aver_{\rm eq}^R \right) } } 
\right] \bs_{{\rm eq}, \, jk}^R \nonumber \\ 
&-& {1 \over 2 \, k_{\rm B} T} \sum_{\mu,\nu = 1 \atop \mu
\ne \nu}^N \, \avel {\bQ}_j  {\bQ}_k E \left( {\br}_{\nu} 
- {\br}_{\mu} \right) \aver_{\rm eq}^R 
\label{covar}
\end{eqnarray}
where, $z^*_D$ is an extension of the definition of the $z^*$
parameter to $D$ dimensions, and angular brackets with superfix ``$R$''
and suffix ``eq'' represent averages carried out with the 
equilibrium distribution function, $\psi_{\rm eq}^R$, 
of the Rouse model, extended to $D$ dimensions. As is well known,
$\psi_{\rm eq}^R$ is a Gaussian distribution function, 
\begin{equation}
\psi_{\rm eq}^R \, ({\bQ}_1, \ldots, {\bQ}_{N-1}) = \left( {H \over 
2 \pi k_{\rm B} T}\right)^{ (N-1) \, {D \over 2}} \!\!\!\!\!\!\!
\exp [ - {1 \over 2} \sum_{j, \, k} {\bQ}_j \cdot 
({\bs}^{- 1})_{{\rm eq}, \, jk}^R
\cdot {\bQ}_k ]
\label{gauss}
\end{equation}
with a covariance matrix, $\bs^R_{{\rm eq}, \, jk}
= \avel {\bQ}_j {\bQ}_k \aver_{\rm eq}^R =  (k_{\rm B} T /H) \,
\delta_{j k} \bu$. Note that $\psi_{\rm eq}$ reduces to $\psi_{\rm
eq}^R$ in the absence of excluded volume interactions. 

The Gaussian nature of the configurational distribution function in
the Rouse model has two consequences, (i) the vector $\br_{\nu \mu}$
between beads $\mu$ and $\nu$ is also a Gaussian distributed random
variable because it is a sum of Gaussian variables, and (ii) the
complex higher order moment in the second term on the right hand side
of Eq.~(\ref{covar}) can be reduced to second order moments by using
general decomposition rules for the moments of a Gaussian
distribution~\citep{prakev2000}. On exploiting these consequences, the
expression for the covariance matrix $\avel {\bQ}_j {\bQ}_k \aver_{\rm
eq}$, correct to first order in $z^*_D$, has the following simple form,
\begin{equation}
\avel \bQ_j \bQ_k \aver_{\rm eq}
= { k_{\rm B} T \over H} \, \left[ \delta_{j k} +  
{1 \over 2 } \,  z_D^*   
\sum_{\mu, \nu =1 \atop \mu \ne \nu}^{N}
{  \theta(\mu,j,k,\nu) \over \left( {d^*}^2 + \vert \mu - \nu \vert 
\right)^{1 + {D \over 2}}} 
\right] \, \bu
\label{covar2}
\end{equation}
The function $\theta(\mu,m,n,\nu)$ has been introduced previously in
the treatment of hydrodynamic interaction~\citep{ottga89}. It is unity if
$m$ and $n$ lie between $\mu$ and $\nu$, and zero otherwise,
\begin{equation}
\theta(\mu,m,n,\nu)=\cases{1& if $\mu \leq m,n < \nu$ \quad or\quad
$\nu \leq m,n < \mu $\cr
\noalign{\vskip3pt}
0& otherwise\cr} 
\label{theta}
\end{equation}
Substituting Eq.~(\ref{covar2}) into Eq.~(\ref{endtoend1}), and carrying out
the sum over the indices $j$ and $k$, one can show
that the mean square end-to-end vector at equilibrium, 
in a space of arbitrary dimensions $D$, correct to
first order in $z_D^*$, is given by,
\begin{equation} 
\avel \br^2 \aver_{\rm eq} =  {D \, k_{\rm B} T \over H} \left[
(N-1)  + {1 \over 2} \, 
z_D^*   \sum_{\mu, \nu =1 \atop \mu \ne \nu}^{N}
{ | \mu - \nu |^2 \over \left( {d^*}^2 
+  | \mu - \nu | \right)^{1 + {D \over 2}}} 
\right]
\label{endtoend2}
\end{equation}

We now consider the limit of a large number of beads, $N$. In this
limit, the sums in Eq.~(\ref{endtoend2}) can be replaced by integrals.
Introducing the following variables,
\begin{equation} 
x = {\mu \over N}; \quad y = {\nu \over N}
\label{ddef}
\end{equation} 
and exploiting the symmetry in $x$ and $y$, we obtain,
\begin{equation} 
\avel \br^2 \aver_{\rm eq} = {D \, k_{\rm B} T \over H} \, N
\left\lbrace 1 +  z^*_D \, N^{\epsilon /2 } \mathop{
\int_{0}^{1} \! \! d x \int_{0}^{x} \! \! d y}_{x > y + c } \, { (x
-y)^2 \over \left( d^2 + x - y \right)^{3 - \epsilon/2}} \right\rbrace
\label{endtoend3}
\end{equation}
where, $\epsilon = 4-D$, $c$ is a {\em cutoff} parameter of order
$1/N$ which accounts for the fact that $\mu \ne \nu$, and the
parameter $d=d^* / \sqrt N$, has already been introduced
earlier. It is worth noting that the leading correction to the
integrals in Eq.~(\ref{endtoend3}) is of order $N^{-1 + { \epsilon /
2} }$~\citep{schaf99}---a fact of particular relevance to the
extrapolation procedure to be adopted in the next section.

From Eq.~(\ref{endtoend3}), it is clear that the excluded volume
corrections to the Rouse end-to-end vector are proportional to
$z^*_D \, N^{\epsilon /2 }$. Therefore, the proper perturbation
parameter to choose is $z_D = z^*_D \, N^{\epsilon 
/2 }$, and not $z^*_D $.  This well known result of
the theory of polymer solutions~\citep{doi86,declos90,schaf99}, indicates
that (i) for $ D =3 $ a perturbation expansion in $z^*_D$ is rendered
useless for long chains, and (ii) useful results can be obtained only
when $D$ is close to 4 dimensions.

The integrals in Eq.~(\ref{endtoend3}) can be
performed analytically, and the resultant expression can be expanded
as a power series in $\epsilon$. For the purposes of renormalization,
the expansion is required only up to order $\epsilon^0$.  Carrying out
the expansion in $\epsilon$, retaining terms up to this order, and
neglecting the cutoff $c$, one obtains,
\begin{equation} 
\avel \br^2 \aver_{\rm eq} =  {D \, k_{\rm B} T \over H} \, N \left\lbrace
1  +  z^*_D \, N^{\epsilon /2 } \, J(d,\epsilon) \, 
\left[ {2 \over \epsilon} - K(d,\epsilon)\right] \, \right\rbrace
\label{endtoend4}
\end{equation}
where, 
\begin{equation} 
 J(d,\epsilon) = (1+3 \, d^2 \, ) \left [ (1+d^2)^{\epsilon/2} 
- d^\epsilon \right] 
\label{jde}
\end{equation}
and,
\begin{equation} 
 K(d,\epsilon) = {1 \over J(d,\epsilon) } \, \left[ {3 \over 2} \,
d^\epsilon \, (1+d^2) +  (1-{3 \over 2} \, d^4) \, (1+d^2 \, )^{-1 +
\epsilon/2} \right] 
\label{kde}
\end{equation}
If we keep $\epsilon$ finite and set $d=0$ in Eq.~(\ref{endtoend4}), we
get the expression for a $\delta$-function excluded volume potential.
This expression has a $(1/\epsilon)$ singularity which arises because
of the neglect of the cutoff. As has been pointed out by
\"Ottinger~\citep{ottrg89}, a renormalization group analysis can be
performed to get rid of the $(1/\epsilon)$ singularity, or, if the
cutoff is retained, renormalization group analysis can be used to get
rid of the cutoff dependence. Curiously, when $d \ne 0$, the parameter
$d$ plays a role similar to the cutoff parameter $c$. This follows
from the fact that Eq.~(\ref{endtoend4}) is free of any singularities as
$\epsilon \to 0$. 

In order to perform the renormalization of the end-to-end vector, we
follow standard practice~\citep{ottrabrg89,ottrg89}, and introduce the
concept of a polymer segment. The polymer segment is used to represent
a unit of molecular weight, free of any mechanical 
interpretation. The polymer chain is thus assumed 
to consist of $N_{\rm s}$ segments, each consisting of
$L_{\rm s}$ beads, so that $N=N_{\rm s} \, L_{\rm s}$. The segments
are introduced in order to remove the ambiguities associated with the
arbitrariness in the choice of the number of beads in a polymer
chain. The size of a polymer segment, $L$, is assumed to be related to
the number of beads in a segment $L_{\rm s}$, through the relation $L=
Z_N \, L_{\rm s}$. Introducing a non-dimensional excluded volume
parameter,
\begin{equation} 
u_{\rm s} =  \left(2 \pi \right)^{D \over 2} \, z^*_D \,  L^{\epsilon
/ 2} \, J(d, \epsilon) 
\label{u0}
\end{equation}
Eq.~(\ref{endtoend4}) can be rewritten in terms of $N_{\rm s}$, $u_{\rm s}$, 
and $L$ as,
\begin{equation} 
\avel \br^2 \aver_{\rm eq} =  {D \, k_{\rm B} T \over H} \, L \, N_{\rm s}
\, Z_N^{-1}   \left\lbrace
1  + {u _{\rm s}\over 4 \pi^2} \, (2 \pi N_{\rm s})^{\epsilon /2 } 
\, Z_N^{- \epsilon /2 } \left[ {2 \over \epsilon} 
- K(d,\epsilon)\right] \, \right\rbrace
\label{endtoend5}
\end{equation}
The next step in the renormalization procedure 
consists of introducing a segment excluded volume parameter $u$, 
which, by renormalizing the {\em bare} excluded volume 
parameter $u_{\rm s}$, takes into account all the excluded volume 
interactions between the many monomers within a segment,
\begin{equation} 
u= Z_u \, u_{\rm s}
\label{renorm}
\end{equation}
The parameters $u_{\rm s}$ and $L_{\rm s}$ are then expected to be functions
of $u$, and expanded in a Taylor's series, 
\[ u_{\rm s} = u ( 1+Au+\ldots) \, ; \quad L_{\rm s}= L(1+Bu+\ldots) \]
so that, for small $u$, 
\begin{equation} 
Z_u = 1 - A u + \cdots \, ; \quad Z_N = 1 - B u + \cdots 
\label{zu}
\end{equation} 
Substituting Eqs.~(\ref{zu}) into Eq.~(\ref{endtoend5}), 
keeping only first order terms in $u$, and expanding 
$(2 \pi N_{\rm s})^{\epsilon /2 }$ to first order in $\epsilon$, 
we have, 
\begin{equation} 
\avel \br^2 \aver_{\rm eq} =  {D k_{\rm B} T \over H} L N_{\rm s}
\left\lbrace 1  + {u \over 2 \pi^2} 
\left( {1 \over \epsilon} + {1 \over 2} \ln (2 \pi  N_{\rm s})
- {K(d,\epsilon) \over 2} + 2 \pi^2  B \right) \right\rbrace
\label{endtoend6}
\end{equation}
Making the choice,
\[ B = - {1 \over 2 \pi^2 \epsilon} +  { K(d,\epsilon) \over 4 \pi^2 } \] 
gets rid of all the {\em micro-structure} dependent terms in
Eq.~(\ref{endtoend6}), and leads to,\footnote{In the limit $d \to 0$,
since $ K(d,\epsilon) $ is a constant equal to one, the quantity $B$
is usually chosen to be equal to $(- 1 / 2 \pi^2 \epsilon)$.  As a
result, an additional term equal to $(- u / 4 \pi^2)$, appears within
the braces on the right hand side of Eq.~(\ref{endtoend7}).}
\begin{equation} 
\avel \br^2 \aver_{\rm eq} =  {D \, k_{\rm B} T \over H} \, L \, N_{\rm s}
\left\lbrace 1  + {u \over 4 \pi^2}  \ln (2 \pi  N_{\rm s}) \right\rbrace
\label{endtoend7}
\end{equation}

The form of Eq.~(\ref{endtoend7}) is not consistent with the 
expectation of a power law  dependence of $\avel \br^2 \aver_{\rm eq}$
on $N_{\rm s}$. Renormalization group analysis resolves this problem
by exponentiating the first order perturbation expansion result,
giving rise to the expression, 
\begin{equation} 
\avel \br^2 \aver_{\rm eq} =  {D \, k_{\rm B} T \over H} \, L 
\left( 2 \pi \right)^{u \over 4 \pi^2}  N_{\rm s}^{2 \, \nu}
\label{endtoend8}
\end{equation}
where, $\nu = {1 \over 2} + {u \over 8 \pi^2}$. 
The exponent $\nu$ is then obtained by substituting the {\em fixed
point} value, $u = u^*$, for $u$~\citep{ottrg89}.  The fixed point
denotes the regime where the properties of the infinitely long polymer
chain have a power law dependence on $N_{\rm s}$, and the fixed point
value is found by setting $Z_u = 0$.  For instance, in the case of a
$\delta$-function excluded volume potential, the fixed point value is
known to be $u^* = (\pi^2 \epsilon / 2)$, which leads to an exponent,
$\nu = {1 \over 2} + {1 \over 16} \, \epsilon$. In the present
instance, the fixed point can be expected to be a function of
$\epsilon$ and $d$.  

It is not possible to find the fixed point value $u^*$ in the present
model by only considering the renormalization of $\avel \br^2
\aver_{\rm eq}$, since the quantity $A$ does not appear in
Eq.~(\ref{endtoend6}), which is correct only to first order in
$u$. However, we do not require the fixed point value in order to make
the following argument.  The form of Eq.~(\ref{endtoend8}) suggests
that, in the limit $N \to \infty$ (which is a consequence of $Z_u =
0$), the parameter $d^*$, which always appears as the combination $d^*
/ \sqrt N$, will not in any way alter the $N_{\rm s}$ dependence of
the mean square end-to-end vector.  Indeed, for any finite value of
$d^*$, the value of $\nu$ will always tend, as $N$ increases, to the
value for a $\delta$-function potential. The validity of the above
renormalization procedure can only be confirmed by ensuring that the
{\em same} definition of $u_{\rm s}$ and choice of parameter $B$ made
here, cancels the micro-structure dependence of all the other
observable quantities in the theory, such as $\eta_p$, $\Psi_1$, etc.
This has not been pursued here as it is considered outside the scope
of the present work. However, the expected independence of the results
from the choice of $d^*$ is in line with the accepted wisdom in static
theories of polymer solutions that the choice of the excluded volume
potential does not influence the scaling of observables with molecular
weight.  Furthermore, dimensionless ratios constructed from observable
quantities are also expected to be free of micro-structure
dependence. In our case, this implies that universal ratios can be
expected to be independent of the choice of the value of $d^*$.

We shall make use of the insight gained in this section in our
analysis of the results of the Gaussian approximation below. This is
justified since the Gaussian approximation and renormalization group
analysis are similar in a certain sense; both are exact to first order
in the strength of excluded volume interaction (the Gaussian
approximaton was shown to be exact to first order in $z^*$ in
\citet{prakev2000}), and both account for an infinite number of higher order
contributions.

\section{The Gaussian approximation in limit of long chains}

\begin{figure}[tbh] \centerline{\epsfbox{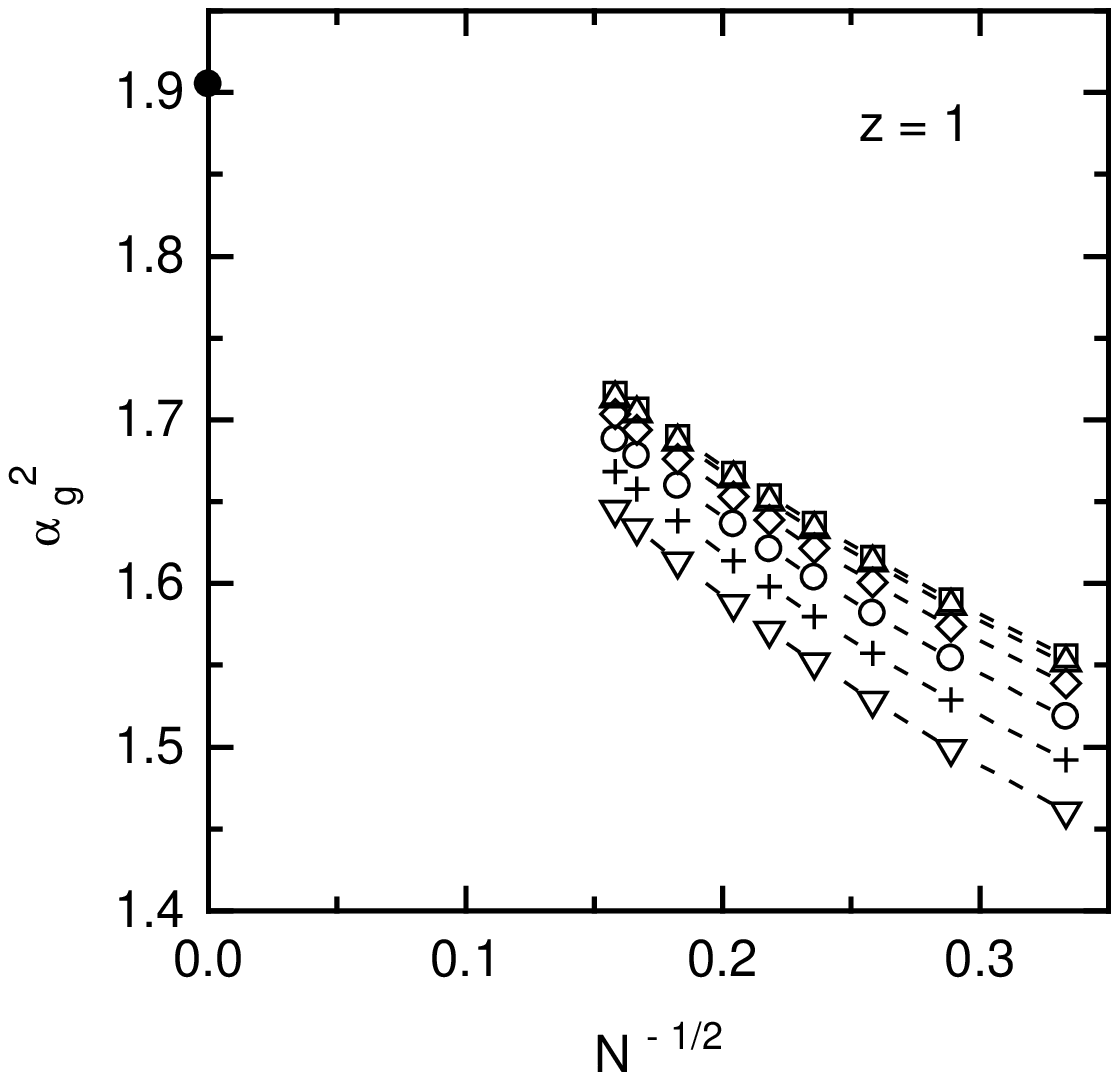}} 
\caption{ \footnotesize Swelling of the radius of gyration 
versus $1 / \sqrt N$, at various values of $d^*$. The symbols
($\square$ : $d^*=0$, $\bigtriangleup$ : $d^*=0.1$,  $\diamondsuit$ :
$d^*=0.2$,  $\bigcirc$ : $d^*=0.3$, $+$ : $d^*=0.4$, 
$\bigtriangledown$ : $d^*=0.5$), are the predictions of the Gaussian 
approximation, while the dashed lines are drawn to guide the eye. The
filled circle on the $y$-axis represents the common extrapolated
value, to $N = \infty$, of all the curves. }
\vskip10pt
\label{fig1} 
\end{figure}

\begin{figure}[tbh] \centerline{ \epsfbox{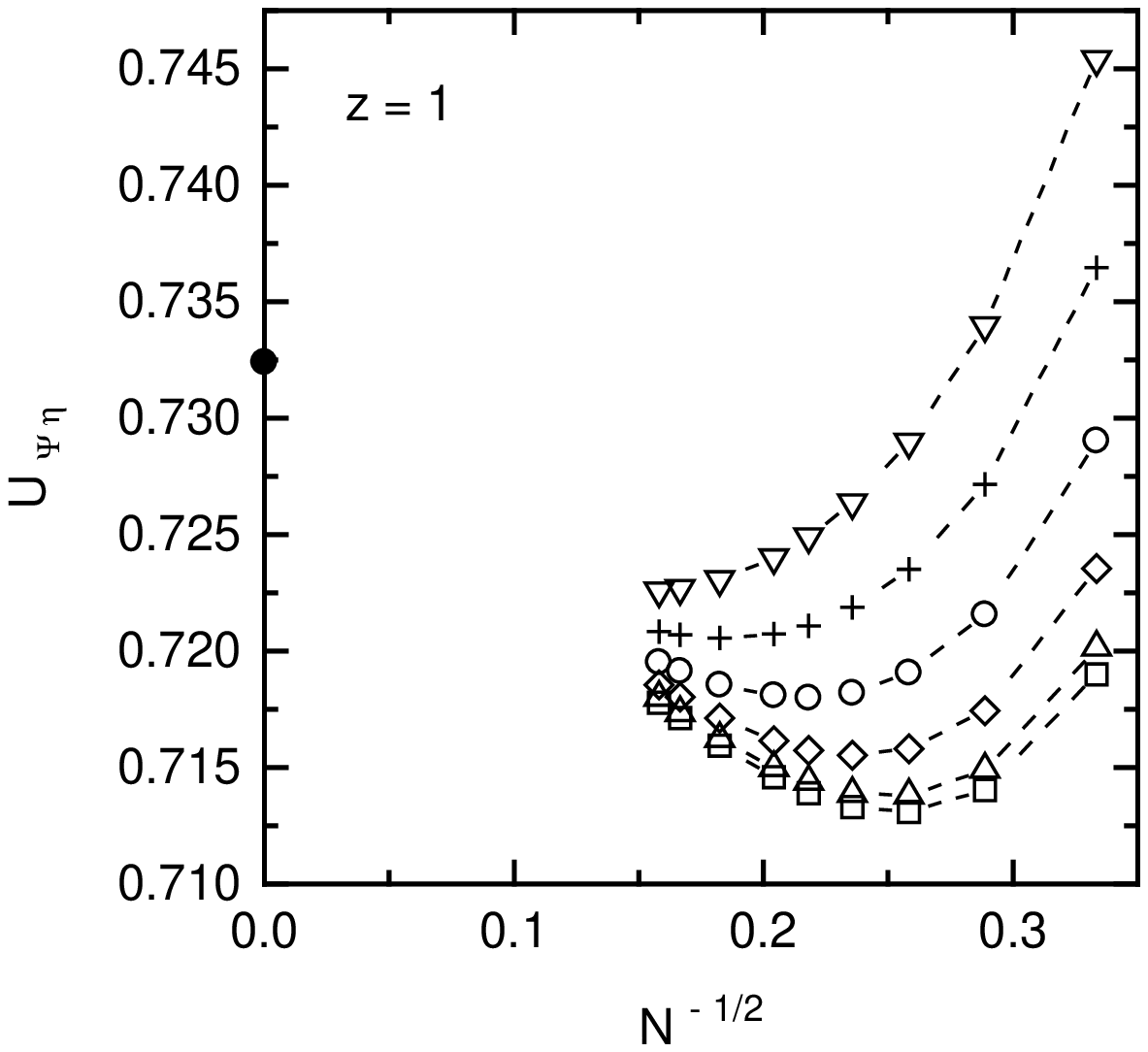}}
\caption{ \footnotesize Non-dimensional ratio constructed from 
the zero shear rate first normal stress difference coefficient 
and the zero shear rate viscosity (see Eq.~(\ref{upsieta}) for
definition), versus $1 / \sqrt N$, at various values of $d^*$. The
symbols ($\square$ : $d^*=0$, $\bigtriangleup$ : $d^*=0.1$,
$\diamondsuit$ : $d^*=0.2$,  $\bigcirc$ : $d^*=0.3$, $+$ : $d^*=0.4$, 
$\bigtriangledown$ : $d^*=0.5$), are the predictions of the Gaussian 
approximation, while the dashed lines are drawn to guide the eye. The
filled circle on the $y$-axis represents the common extrapolated
value, to $N = \infty$, of all the curves. }
\vskip10pt
\label{fig2} 
\end{figure}

\begin{table}[bt]
\caption{Asymptotic values of ratios of equilibrium and zero shear
rate properties, at $z=1$. Numbers in parentheses indicate the
uncertainity in the last figure.} 
\label{tab1}
\begin{center}
\begin{tabular}{ccccccccc}	
 & & & & & & & &   \\ \hline
 & & & & & & & & \\
 $\alpha^2$ & & $\alpha^2_g$ & & $(\Psi_{1,0} / \Psi_{1,0}^R)$ & & 
 $U_{\Psi \eta }$ & & $U_{R}$  \\
 & & & & & & & &   \\ \hline
 & & & & & & & & \\ 
1.960 (3)   & & 1.905 (4) & & 3.32 (1) & & 0.7324 (4) & & 0.9723 (3) \\
 & & & & & & & &   \\ \hline
\end{tabular}
\end{center}
\vskip20pt
\end{table}

We now examine the aymptotic predictions of the Gaussian approximation,
in order to examine the role of the parameter $d^*$. The
renormalization group arguments of the previous section
indicate that, for large values of the number of
beads $N$, one would expect the scaling with $N$, of various
observable quantities, to become independent of 
$d^*$. A similar behavior is expected of non-dimensional ratios
constructed with these quantities.

Figures~{\ref{fig1}} and~{\ref{fig2} examine the dependence of two
quantities, predicted by the Gaussian approximation, on $1 / \sqrt N$,  
for various values of $d^*$, at $z =1$. The first quantity,
$\alpha^2_g$, is both 
an equilibrium property and a linear viscoelastic property, since (as
was shown in~\citet{prakev2000}), it describes the equilibrium
swelling of the radius of gyration, and the ratio of the zero shear
rate viscosity in the presence of excluded volume interactions to the
zero shear rate viscosity in the Rouse model,
\begin{equation}
\alpha^2_g = {\avel R_g^2 \, \aver_{\rm eq} \over 
\avel R_g^2 \, \aver_{\rm eq}^R} = {\eta_{p,0} \over \eta_{p,0}^R}
\label{alphag}
\end{equation}
Here, $\avel R_g^2 \, \aver_{\rm eq}$ is the radius of gyration, and
a superfix ``$R$'' on a quantity indicates the Rouse model value of
the quantity. The second quantity is a non-dimensional ratio, $U_{\Psi
\eta }$, defined by,
\begin{equation} 
U_{\Psi \eta } = {{n_{\rm p} k_{\rm B} 
T \Psi_{1, 0} \over { \eta_{p, 0}^2}}}
\label{upsieta}
\end{equation} 
The dependence of these quantities on $1 /\sqrt N$ was examined for 
two reasons, (i) the leading order correction, in 3 dimensions, to the
integrals in Eq.~(\ref{endtoend3}) is of order $1 /\sqrt N$,
and (ii) the renormalisation group calculation of the previous
section, and asymptotic results obtained earlier with the Gaussian
approximation~\citep{prakev2000}), 
suggest that the parameter $d^*$ always appears in the theory as the
re-scaled variable $d = d^* / \sqrt N$. As a result, the leading order
corrections to the infinite chain length limit, of all material
properties, are expected to be functions of $1 /\sqrt N$.
In the limit $N \to \infty$, therefore, all material properties should
become independent of $d^*$, and depend only on the variable $z$. 

The filled circles on the $y$-axis in Figs.~{\ref{fig1}}
and~{\ref{fig2}, represent the {\em common} extrapolated value, to $N
= \infty$, of each of the curves for the various values of $d^*$. The
values for different $N$, for each value $d^*$, were extrapolated to
the limit $N \to \infty$, using a rational function extrapolation
algorithm~\citep{numrec92}. Clearly, both $\alpha^2_g$ and $U_{\Psi
\eta }$ become independent of $d^*$ in the limit $N \to \infty$.

The variables, (i) $\alpha^2 = {\avel \br^2 \aver_{\rm eq} / \avel
\br^2 \aver_{\rm  eq}^R }$, which is the swelling of the end-to-end
vector at equilibrium, (ii) $(\Psi_{1,0}/ \Psi_{1,0}^R)$, which is 
the ratio of the zero shear rate first normal stress difference in the
presence of excluded volume interactions to the zero shear rate first
normal stress difference in the Rouse model, and (iii) $U_{R}$, which
is a non-dimensional ratio of the radius of gyration to the end-to-end
vector at equilibrium, defined by,   
\begin{equation} 
 U_{R} = 6 \, 
{\avel R_g^2 \, \aver_{\rm eq} \over  
\avel \br^2 \, \aver_{\rm eq} } 
\label{uR}
\end{equation} 
exhibit the same behavior (as that displayed in Figs.~{\ref{fig1}}
and~{\ref{fig2}), in the limit $N \to \infty$. The asymptotic
values, for $z=1$, of each of these variables is given in
Table~{\ref{tab1}}.\footnote{The non-dimensional
ratio, $U_{\eta R} =  \left[ \eta_{p, 0} / n_{\rm p} \eta_s (4 \pi /3
) \avel R_g^2 \, \aver_{\rm eq}^{3/2} \right]$, is not a 
universal ratio in the present model since it scales with $N$ as 
$N^{1 -\nu}$. It becomes a universal ratio only when hydrodynamic 
interaction effects are included in the model.}

\begin{figure}[tbh] \centerline{\epsfbox{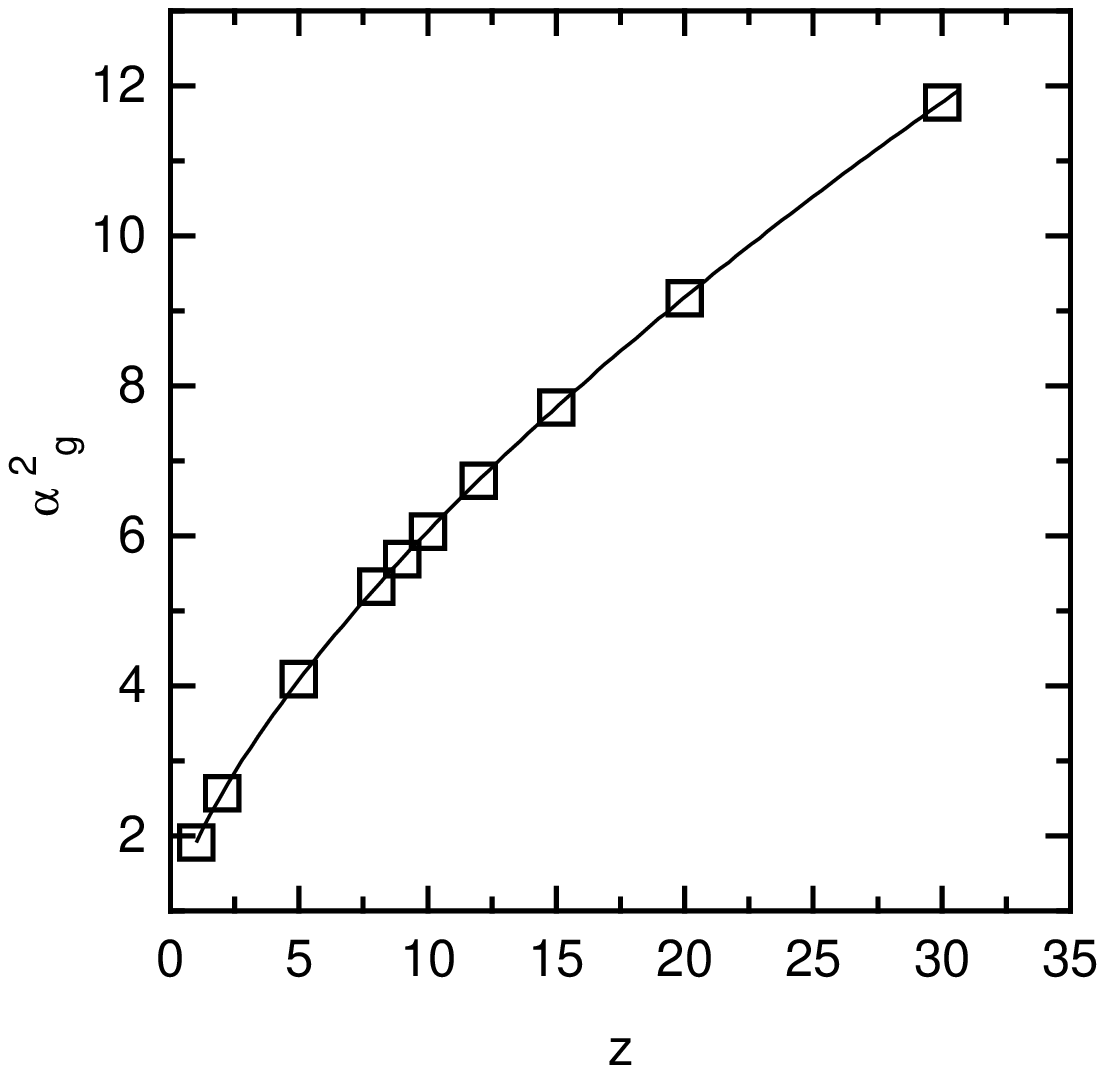}} 
\caption{ \footnotesize Asymptotic swelling of the radius of gyration
versus $z$. The symbols are the results of the Gaussian approximation,
while the line is a curve fit using an equation of the
form given in Eq.~(\ref{curfit}). The curve fit parameters are given in
Table~\ref{tab2}. }
\vskip10pt
\label{fig3} 
\end{figure}

\begin{figure}[tbh] \centerline{ \epsfbox{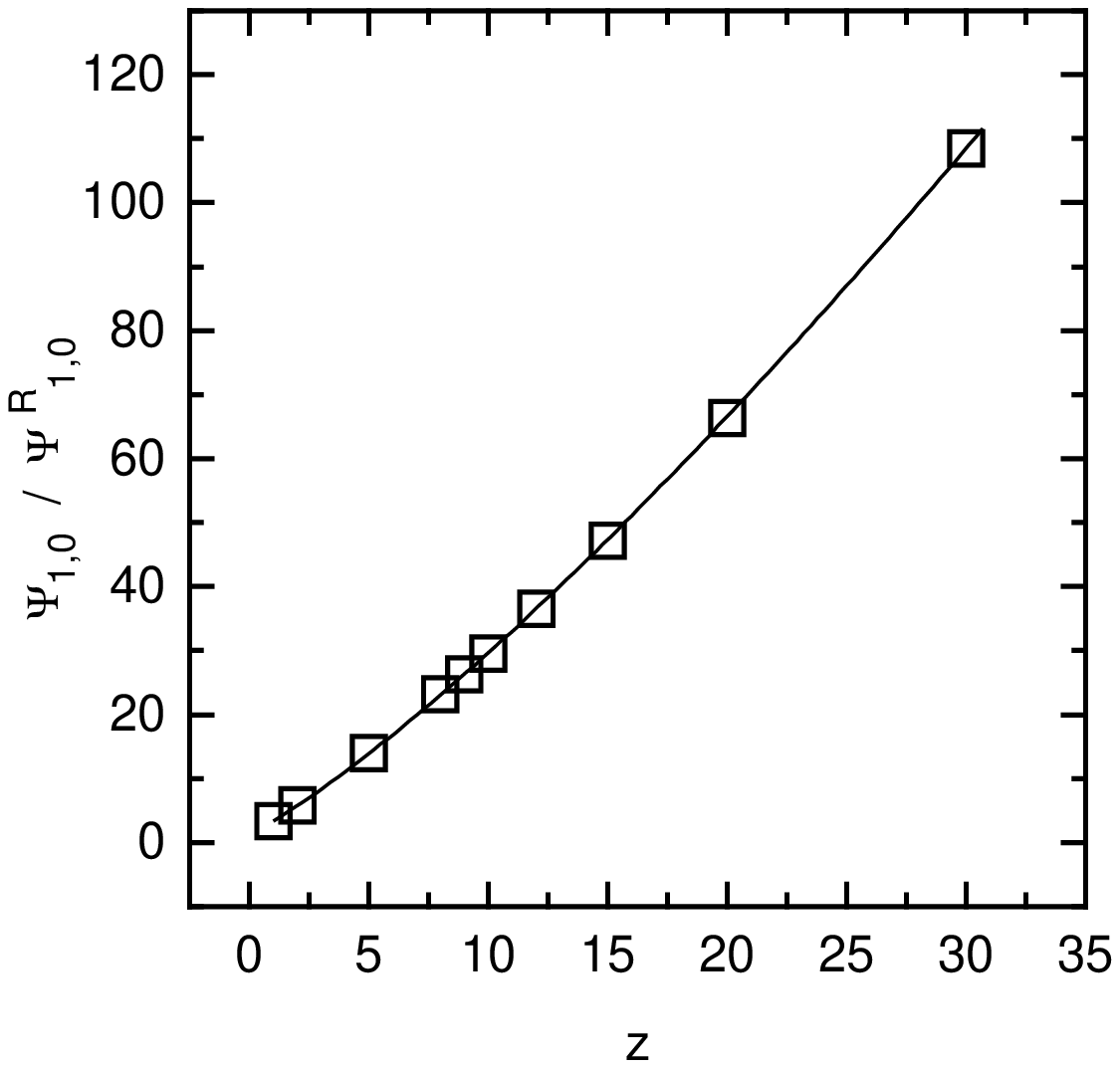}}
\caption{ \footnotesize Asymptotic ratio of the zero shear rate first
normal stress difference coefficient in the presence of excluded
volume interactions to the zero shear rate first normal stress
difference coefficient in the Rouse model versus $z$. The symbols are
the results of the Gaussian approximation, while the line is a curve
fit using an equation of the form given in Eq.~(\ref{curfit}). The
curve fit parameters are given in Table~\ref{tab2}. }
\vskip10pt
\label{fig4} 
\end{figure}

\begin{table}[bt]
\caption{Parameters appearing in Eq.~(\ref{curfit}), used 
to fit the asymptotic predictions of the Gaussian approximation,
displayed in Figs.~{\ref{fig3}} and~{\ref{fig4}}.}
\label{tab2}
\begin{center}
\begin{tabular}{ccccccc}	
 & & & & & &  \\ \hline
 & & & & & &  \\
 & & $\alpha^2$ & & $\alpha^2_g$ & & $(\Psi_{1,0} / \Psi_{1,0}^R)$  \\
 & & & & & &  \\
 & & & & & &  \\ \hline
 & & & & & &  \\
$a$ & & 4.352 & & 4.178 & & 4.058 \\
 & & & & & &  \\
$b$ & & 2.715 & & 2.453 & & 1.755 \\
 & & & & & &  \\
$m$ & & 0.323 & & 0.318 & & 0.630 \\
 & & & & & &  \\ \hline
\end{tabular}
\end{center}
\vskip10pt
\end{table}

\begin{figure}[tbh] \centerline{ \epsfbox{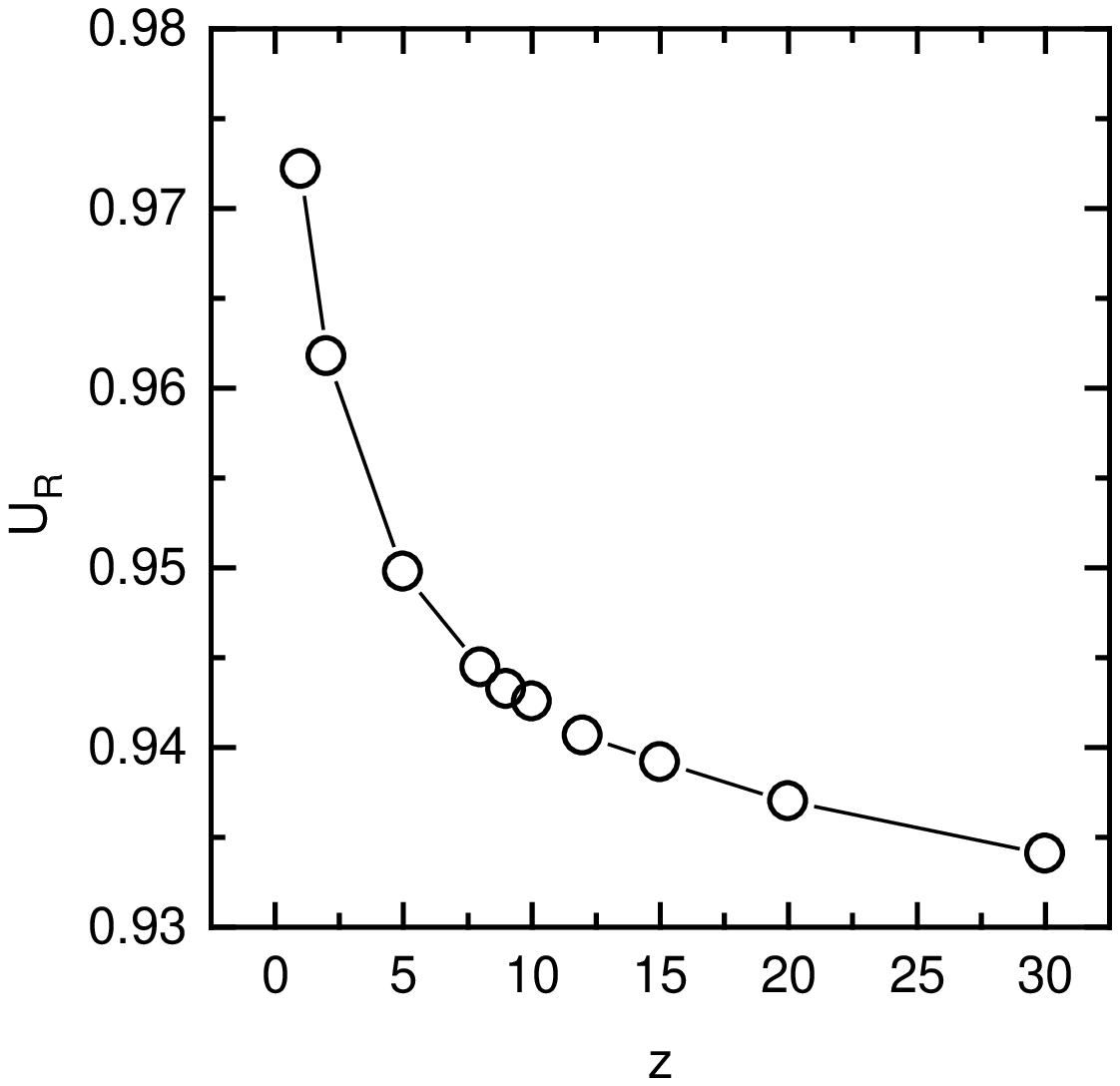}}
\caption{ \footnotesize Universal ratio constructed from 
the radius of gyration and the end-to-end vector versus $z$. 
The symbols are the results of the Gaussian
approximation, while the line is drawn to guide the eye.}
\label{fig5} 
\vskip10pt
\end{figure}

\begin{figure}[tbh] \centerline{ \epsfbox{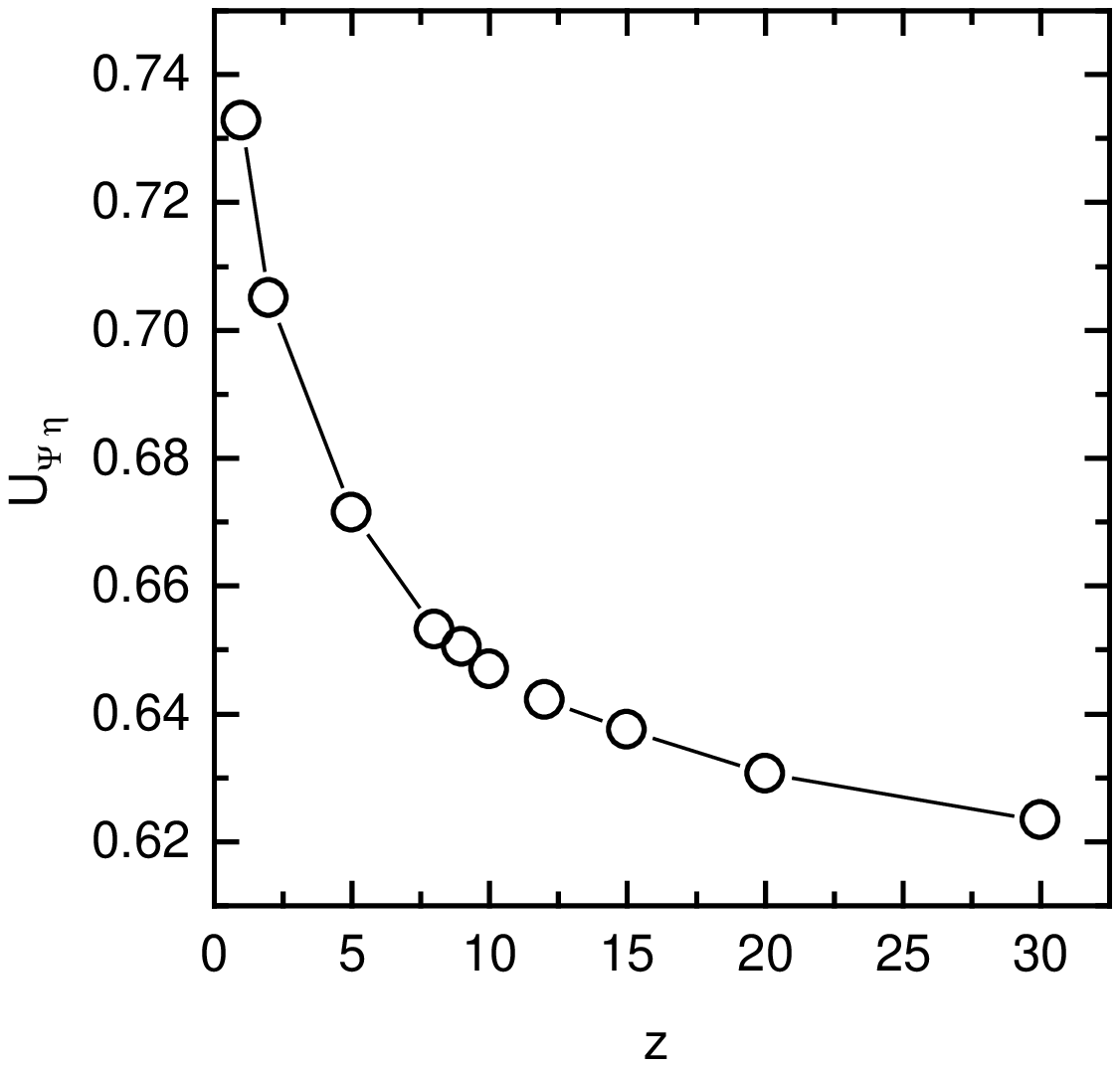}}
\caption{ \footnotesize Universal ratio constructed from 
the zero shear rate first normal stress difference coefficient 
and the zero shear rate viscosity versus $z$. The symbols are the
results of the Gaussian approximation, while the line is drawn to
guide the eye.  } 
\label{fig6} 
\vskip10pt
\end{figure}

Figures~{\ref{fig3}} and~{\ref{fig4} describe the {\em asymptotic\/}
$z$ dependence of $\alpha^2_g$ and $(\Psi_{1,0}/ \Psi_{1,0}^R)$. The
behavior of $\alpha^2$ has not been displayed as it is very similar to 
that of $\alpha^2_g$.  The symbols in Figs.~{\ref{fig3}}
and~{\ref{fig4} are the asymptotic predictions of the Gaussian
approximation. These predictions were obtained, at each value of $z$,
by using the extrapolation procedure described above for $z=1$. 

The lines through the symbols are curve fits, using an equation which
has a form commonly used to fit results of renormalization group
analysis~\citep{schaf99},
\begin{equation} 
y = \left( 1 + a \, z + b \, z^2  \right)^m
\label{curfit}
\end{equation} 
where, $y$ represents the fitted variable. The parameters 
$a$, $b$ and $m$, for $\alpha^2$, $\alpha^2_g$ and 
$(\Psi_{1,0}/ \Psi_{1,0}^R)$,  are given in Table~\ref{tab2}.
The maximum difference between the computed data and the 
curve fit for all the three properties was less than 0.5\%. 

The $N$ dependence of the equilibrium properties $\avel \br^2 \,
\aver_{\rm eq}$ and $\avel R_g^2 \, \aver_{\rm eq}$, for large values
of $N$, can be obtained from Eq.~(\ref{curfit}), and the definitions
of the quantities $\alpha^2$ and $\alpha^2_g$. Both the properties
scale identically with molecular weight, namely, as $N^{2 \, \nu}$.
Experimental results suggest a value of $\nu = 0.592 \pm
0.003$~\citep{hay99}, while renormalization group calculations and
Monte Carlo simulations suggest a value of $\nu =
0.588$~\citep{schaf99,gras99}.  Since $z$ scales as $N^{1/2}$, it is
clear from Eq.~(\ref{curfit}) that for large values of $N$,
$\nu=(1+m)/2$.  From Table~{\ref{tab2}}, one can see that the values
of $m$ for $\alpha^2$ and $\alpha^2_g$, imply a Gaussian approximation
prediction of $\nu$ between 0.659 and 0.662. We expect the value of
$m$ for $(\Psi_{1,0}/ \Psi_{1,0}^R)$ to be twice the value of $m$ for
$\alpha^2_g$. This follows from Eq.~(\ref{alphag}), since $\Psi_{1,0}$
scales as $\eta_{p,0}^2$, and $\eta_{p,0}^R$ and $\Psi_{1,0}^R$ scale
with $N$ as $N^2$ and $N^4$, respectively. As can be seen from
Table~{\ref{tab2}}, this expectation is reasonably fulfilled.

It must be noted that the present data, which has been accumulated
for relatively small values of $z$, might still be describing the
crossover region between Rouse scaling and the final scaling in the
`excluded volume limit' of large $z$. In that case, the curve fit
parameter $m$ might be modified as data is compiled for larger values
of $z$. In calculations based on renormalization group arguments, the
convergence to a value $\nu = 0.588 \pm 0.01$ is relatively fast
(roughly by $z=18$), but the final convergence is very slow (occuring
at $z > 100$)~\citep{schaf99}.

The problem with going to larger values of $z$ with the Gaussian
approximation is that, in order to maintain the accuracy of the
numerical extrapolation procedure, one must have data for values of $N
> 40$, which, as was mentioned earlier, requires large amounts of CPU
time. In the treatment of the non-linear microscopic phenomenon of
hydrodynamic interaction, the use of a normal mode approximation,
simultaneously with the Gaussian approximation, led to a
non-perturbive solution scheme which was as accurate as the 
Gaussian approximation, but significantly
less computationally intensive. As a result, chains with lengths up to
$N=100$ could be examined~\citep{prakotthi97}. It is worth
examining if a normal mode approximation in the present situation also
leads to a significant reduction in the computational intensity.

Figures~{\ref{fig5}} and~{\ref{fig6} examine the asymptotic $z$
dependence of the two universal ratios, $U_R$ and $U_{\Psi \eta
}$, predicted by the Gaussian approximation. 
In the Rouse model, $U_{R} = 1$. Both renormalization 
group calculations and Monte Carlo simulations yield an 
identical value of $U_{R} = 0.959$ in the excluded volume 
limit~\citep{schaf99}. Though the curve in Fig.~{\ref{fig5}}
decreases relatively rapidly for small values of $z$, and
appears to be levelling off as $z$ increases, it still has a
non-zero slope at $z=30$. This suggests that the asymptotic 
value of the ratio, at large values of $z$, is yet to be 
reached, and that the present data is still in the crossover region. 

For long chains, $U_{\Psi \eta} = 0.8$ in the Rouse model. On the
other hand, a renormalization group calculation yields a value
$U_{\Psi \eta } = 0.6288$~\citep{ottrg89}. The value of $U_{\Psi
\eta }$ in Fig.~{\ref{fig6}}, at $z=30$, is quite close to the
prediction of the renormalization group calculation (which was based
on a $\delta$-function potential). However, the non-zero slope at $z=30$
indicates that the excluded volume limit is yet be reached. 

\section{Conclusions}

Renormalisation group arguments, carried out in the context of the
end-to-end vector at equilibrium, seem to suggest that the scaling
with molecular weight of the equilibrium and zero shear rate
properties, and the values of non-dimensional ratios of these
quantities, should become independent of $d^*$ in the limit of large
$N$. The asymptotic results of the Gaussian approximation have been
examined in the light of these expectations. The data clearly
indicates independence from the choice of $d^*$. The aymptotic
dependence on $z$ presented here, of the universal ratios $U_{\Psi
\eta}$ and $U_R$, appears to lie in the crossover region between Rouse
scaling and the scaling in the excluded volume limit of large $z$.
Accumulation of more data at larger values of $z$ is required before
the scaling of the Gaussian approximation in the excluded volume limit
can be described unambiguously. This task is rendered difficult
because of the computational intensity of the Gaussian approximation,
which scales with chain length as $N^6$.

\begin{ack} 
Support for this work through a grant III. 5(5)/98-ET from the
Department of Science and Technology, India, is acknowledged. Most of
this work was carried out while the author was an Alexander von
Humboldt fellow at the Department of Mathematics, University of
Kaiserslautern, Germany. The High Performance Computational Facility
at the University of Kaiserslautern is gratefully acknowledged for
providing the use of their computers.
\end{ack}

\end{document}